# Preparation and characterization of an anionic dye-polycation molecular films by electrostatic Layer-by-Layer adsorption process

D. Dey[a], S. A. Hussain[a], R. K. Nath[b] and D. Bhattacharjee[a]

[a]Department of Physics, Tripura University, Suryamaninagar-799130, Tripura, INDIA
[b]Department of Chemistry, Tripura University, Suryamaninagar-799130, Tripura, INDIA

**Abstract:**

This communication reports the formation and characterization of self assembled films of a low molecular weight anionic dye amaranth and polycation Poly (allylamine hydrochloride) (PAH) by electrostatic alternating Layer-by-Layer (LBL) adsorption. It was observed that there was almost no material loss occurred during adsorption process. The UV-Vis absorption and fluorescence spectra of amaranth solution reveal that with the increase in amaranth concentration in solution, the aggregated species starts to dominate over the monomeric species. New aggregated band at 600 nm was observed in amaranth-PAH mixture solution absorption spectrum. A new broad low intense band at the longer wavelength region, in the amaranth-PAH mixture solution fluorescence spectrum was observed due to the closer association of amaranth molecule while tagged into the polymer backbone of PAH and consequent formation of aggregates. The broad band system in the 650-750 nm region in the fluorescence spectra of different layered LBL films changes in intensity distribution among various bands within itself, with changing layer number and at 10 bilayer LBL films the longer wavelength band at 710 nm becomes prominent. Existence of dimeric or higher order n-meric species in the LBL films was confirmed by excitation spectroscopic studies. Almost 45 minute was required to complete the interaction between amaranth and PAH molecules in the 1-bilayer LBL film.

E mail: sa_h153@hotmail.com

## 1. Introduction:

Layer-by-Layer (LBL) self assembled technique is the process of obtaining thin organic films which forms spontaneously on solid surfaces [1]. They are the subject of intense study because of their potential utility in such applications as wetting, adhesion, lubrication, high resolution lithography [1-3], molecular electronic devices [4,5], electroluminescent devices [5,7] and second harmonic generation [8]. The technique was originally developed for the sequential adsorption of oppositely charged polyelectrolytes on solid substrates [1,2]. However, recent studies show that the sequential adsorption approach can be used to manipulate many different types of materials including conducting polymers [9], light emitting materials [10], nonlinear optical polymers [11], inorganic nanoparticles [12], biomaterials [13], dyes [14] and various other organic, inorganic and polymeric systems [15,16]. This enables the arrangement of materials at a molecular level to produce cooperative electronic and optical properties.

Since the LBL self assembled technique relies on the electrostatic interaction of complementary anion and cation pair in successive adsorption steps, polyions are generally used for LBL film deposition. However the current research interests lies on the fabrication of LBL films using low molecular weight organic materials. Since these organic materials offer many interesting electronic and optical properties.

Most of the small organic molecules cannot usually be manipulated into molecular layers by LBL technique because of the fact that, due to the presence of small number of charged groups in these molecules, physiadsorption of such molecules in LBL films is not straight forward. Material loss by washing is substantial in many cases. There are few reports of LBL self assembled films of organic molecules. However, in most of the cases the molecules had long alkyl chain [17] or a correct combination of polyions was used. Moreover in most of the cases material loss by washing was prominent [18].

In the present communication we report the successful incorporation and detail photophysical studies of a low molecular weight biologically important anionic dye amaranth [19], into the LBL films alongwith the poly (allylamine hydrochloride) (PAH). The most interesting thing in our observation was that no material loss by washing of LBL films was observed. The simplicity of this deposition mode suggest that the LBL film deposition technique can now be applied to any suitable low molecular weight organic molecules with few charge groups present, in combination with a suitable polyanion or polycation, without any material loss or release during the subsequent layers physiadsorption.

The anionic dye amaranth used in this study is a product from amaranth plant which is a dicotyledoneous plant with well-balanced grain proteins and has been proposed as a new alternative source of good quality proteins [20]. Amaranth can be used for natural and synthetic fibers, leather, paper and phenol formaldehyde resins. It is also used as food dye and to colour cosmetics [19].

Characterizations of amaranth-PAH LBL self assembled films have been done by using UV-Vis absorption and fluorescence spectroscopic techniques.

## 2. Experimental:

The anionic dye used in this work is amaranth. Poly (allylamine hydrochloride) (PAH) is used as the polycation. Both the dye amaranth (molecular weight = 604.47), purity > 99% and the PAH (molecular weight = 70,000), purity > 99%, were purchased from Aldrich Chemical Co.

and were used without any further purification. The chemical structure of both the dye and the PAH are shown in figure 1. The electrolyte deposition bath was prepared with $10^{-3}$ M (based on the repeat units for polyion) aqueous solutions using triple distilled deionised (18.2 MΩ) Millipore water.

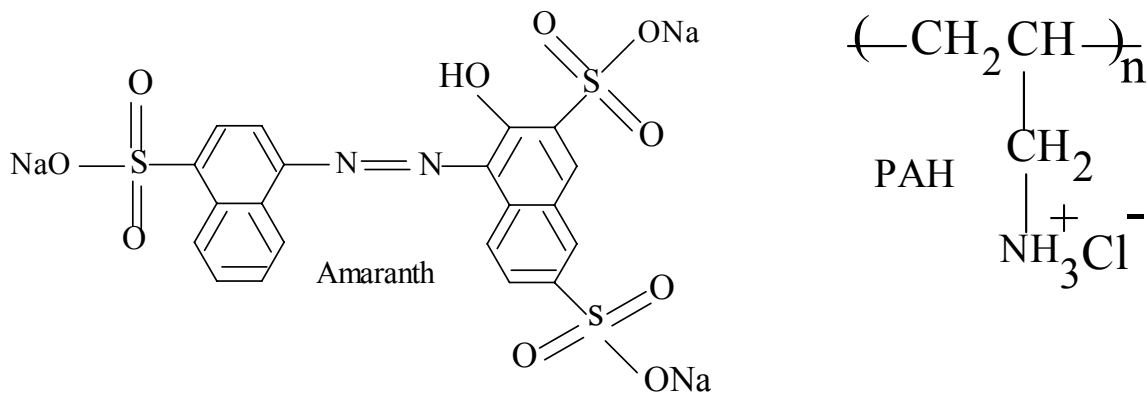

*Fig. 1*: *Molecular structure of amaranth and Poly (allylamine hydrochloride) (PAH) molecule.*

Layer-by-layer (LBL) self-assembled films were obtained by dipping thoroughly cleaned fluorescence grade quartz substrates alternately in solutions of the dye and the oppositely charged polyelectrolyte PAH. LBL method utilizes the Vanderwaals interactions between the quartz substrate and PAH as well as charge transfer (CT) interaction between PAH and the dye amaranth, which contains anionic groups and therefore used as anion. PAH was used as polycation for the fixation of the dye molecule to the substrate. First of all the quartz substrate was cleaned by standard procedure [2] and immersed in the PAH solution for 15 minutes followed by same rinsing in water bath for 2 minutes. The rinsing washes off the surplus cation attached to the surface. The substrate was then immersed in amaranth solution for 15 minutes followed by same rinsing procedure. After each deposition and rising procedure sufficient time was allowed to dry up the film and their UV-Vis absorption spectra were recorded to monitor the film growth. Deposition of the PAH (cation) and amaranth (anion) layers resulted in one bilayer of self-assembled film. The whole sequence of the film deposition procedure was repeated for the preparation of desired number of bilayer LBL films.

The UV-Vis absorption and steady state fluorescence spectra of the LBL films as well as solutions were recorded using Lambda-25 UV-Vis spectrophotometer, Perkin Elmer and LS-55 fluorescence spectrophotometer, Perkin Elmer respectively.

### 3. Results and Discussions:

Figures 2a and 2b show the UV-Vis absorption and steady state fluorescence spectra of the aqueous solution of amaranth at different concentrations alongwith the solution spectra of amaranth-PAH mixture as well as the pure amaranth microcrystal spectrum for comparison.

The solution absorption spectra (figure 2a) show distinct and intense band systems in the 200-650 nm region with an intense and sharp monomeric band having peak at 210 nm and weak humps at 246 and 288 nm alongwith a broad J-aggregate band at around 525 nm. The amaranth solution absorption spectra for different concentrations ($10^{-5} - 10^{-8}$M) are almost similar except changes in intensity distribution. From the plot (left inset of figure 2a) of the intensity of the monomeric band at 210 nm and the aggregate band at 525 nm, as a function of the concentration of the dye solution it was observed that the intensity of both the bands increases with the increase in dye concentration in the solution. The right inset of figure 2a shows that with the increase in amaranth concentration in solution, the absorption band at 525 nm due to aggregated species starts to dominate as compared to the monomeric band at 210 nm. This indicates that closer association of amaranth molecules takes place with increasing concentration in solution.

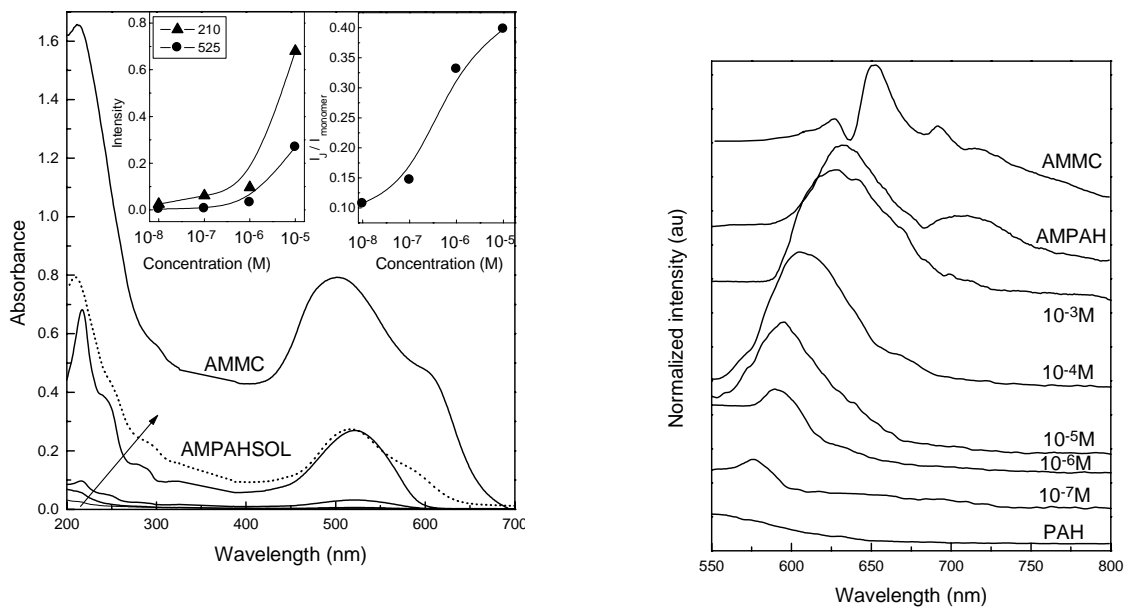

*Figure 2a*: Absorption spectra of amaranth ($10^{-8} – 10^{-5}$ M) aqueous solution alongwith the amaranth microcrystal and amaranth-PAH (1:1 volume ratio) mixture solution spectrum. The left inset shows the plot of intensity of 210 nm and 525 nm band as a function of concentration. The right inset shows the concentration variation of the ratio between the intensities of the 525 nm and the 210 nm absorption bands.

*Figure 2b*: Fluorescence spectra of amaranth ($10^{-7} – 10^{-3}$ M) aqueous solution alongwith the amaranth-PAH (1:1 volume ratio) mixture (AMPAH), pure PAH solution spectrum (PAH) and amaranth microcrystal spectrum (AMMC).

Amaranth microcrystal absorption spectrum also shows (figure 2a) similar band pattern as that of the pure amaranth solution absorption spectrum except a weak hump in the longer wavelength region with peak at 600 nm is observed. The origin of this band is due to the closer association and aggregates of amaranth molecules in the microcrystal. This aggregation band is totally absent in pure solution absorption spectrum of amaranth even at higher concentration.

The amaranth-PAH mixture solution (1:1 volume ratio) absorption spectrum is also shown in figure 2a. A weak hump at around 600 nm appears in the absorption spectrum of amaranth-PAH mixture solution, along with all other bands which are present in the pure amaranth solution absorption spectrum. This 600 nm band was absent in the pure amaranth solution absorption spectrum. The origin of this band is not readily explicable. However, a comparison with the microcrystal absorption spectrum of pure amaranth clearly indicates that this new band is due to the aggregated species of amaranth in the mixed solution.

A closer look at the interaction scheme between PAH and amaranth molecule (as shown in figure 3) reveals that the anionic part of the amaranth molecule interacts with the cationic part of the PAH molecule. Thus in the complex species, the amaranth molecules get closer side by side and closer association of amaranth molecule takes place, which is manifested by the emergence of the new hump at the longer wavelength region of amaranth-PAH mixture solution spectrum.

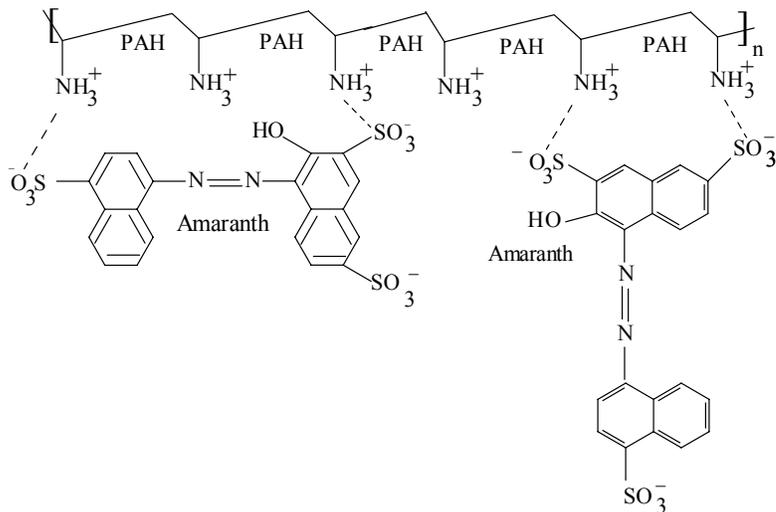

*Fig. 3: Schematic representation of amaranth-PAH interaction scheme.*

The amaranth solution fluorescence spectra (figure 2b) show distinct band system within 550-700 nm region with peak position varying at around 585-630 nm for different concentrations ($10^{-3} - 10^{-7}$M) of the amaranth in solution. The fluorescence band maxima increases in intensity with the increase in amaranth concentration and also red shifted. The increase in intensity with the increase in concentration and red sift may be due to the predominance of aggregated species with the increase in amaranth concentration.

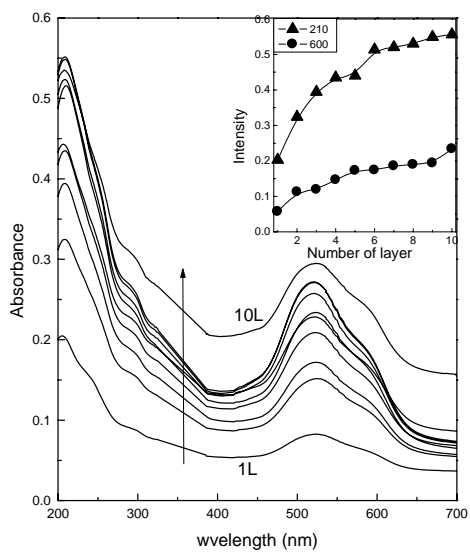

*Figure 4a: Absorption spectra of different layered (1-10 bilayer) amaranth-PAH layer-by-layer self assembled films. Inset shows the plot of intensities of 210 nm and 600 nm band as a function of layer number.*

Along with the pure amaranth solution band, a new broad band in the longer wavelength region with peak at around 710 nm in the amaranth-PAH mixture solution fluorescence spectrum was observed, indicating the interaction between the dye and the polycation PAH molecule. Amaranth microcrystal fluorescence spectrum (also shown in figure 2b) also shows a low intense broad band in the longer wavelength region indicating closer association of amaranth molecules. Therefore it can certainly be concluded that the broad band in the longer wavelength region in the amaranth-PAH mixture solution fluorescence spectrum originates due to the closer association of amaranth molecules on the backbone of PAH molecules.

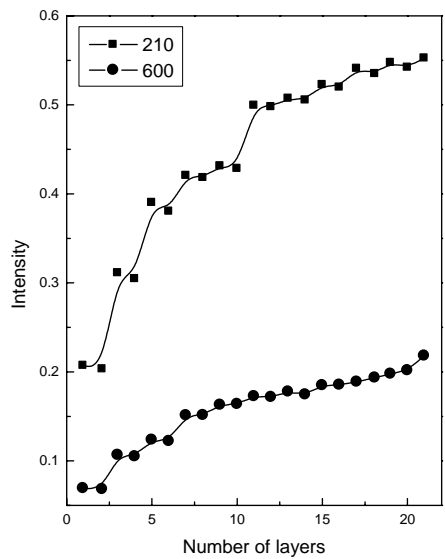
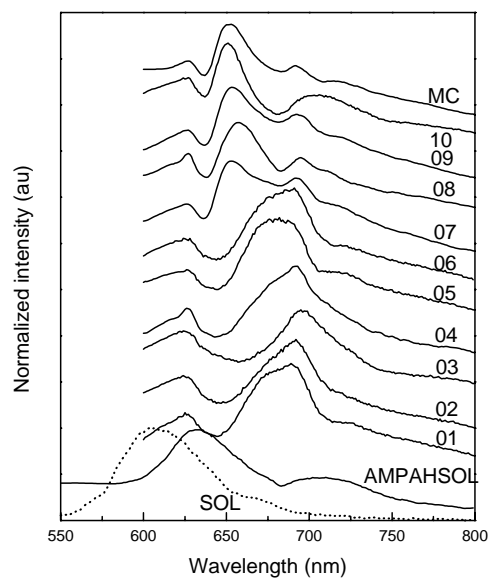

*Figure 4b*: Plot of intensities of 210 nm and 600 nm band after each monolayer deposition of either amaranth or PAH as a function of layer number.

*Figure 4c*: Fluorescence spectra of different layered (1-10 bilayer) amaranth-PAH layer-by-layer self assembled films alongwith the microcrystal (MC), pure amaranth solution (SOL) and amaranth-PAH mixture solution (AMPAHSOL).

Figure 4a shows the UV-Vis absorption spectra of different layered (1-10 bilayer) PAH-amaranth self-assembled LBL films. It is interesting to note that the absorption spectra of different layered LBL films show almost similar band pattern irrespective of the layer number except an increase in intensity and have distinct similarity and identical band position with that of the amaranth microcrystal absorption spectrum. This is a clear indication that closer association of amaranth molecules occur in the Amaranth-PAH LBL films due to the interaction with the PAH molecules.

The inset of figure 4a shows the plot of intensity of the monomeric band (210 nm) and the 600 nm band as a function of layer number. From the figure it was observed that the intensity of both the band increases systematically with the increase in layer number. The increase in intensity of the 600 nm band in the absorption spectra of the LBL films with the increase in layer number indicates the increase in amaranth-PAH complex species with the increase in film thickness and definitely confirms the successful incorporation of amaranth molecules in the PAH-amaranth LBL films.

To monitor the growth of the films as well as to check whether there is any material loss during film deposition, the absorption spectra were recorded after deposition of each layer. Figure 4b shows the intensity of absorption maxima for 210 nm and the 600 nm band as a function of number of deposited layers. The first absorption spectrum was recorded after the deposition of amaranth layer and then the absorption spectra were recorded after deposition of each layer either the polycation (PAH) or the dye (amaranth). The deposition was started with the polycation deposition and ended up with the dye deposition. Figure 4b represents 21 layer or 11 bilayer of self assembled PAH-amaranth LBL films. A closer look at the figure reveals that the intensities of the absorption maxima increases after deposition of each amaranth layer and remained almost constant after each PAH layer deposition but does not decrease in intensity. This confirms that the amaranth molecules do not come out of the film during PAH deposition. Although in few other cases the material loss was reported during the polycation deposition [18]. But here the PAH-amaranth LBL films could be prepared almost without any material loss.

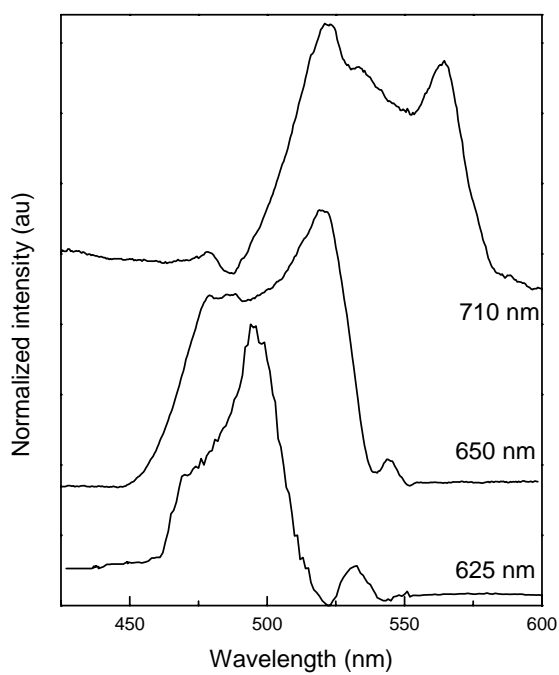

*Figure 4d:* *Excitation spectra of 10 bilayered layer-by-layer self assembled film with monitoring wavelengths 625, 650 and 710 nm.*

Figure 4c represents the normalized fluorescence spectra of different layered (1-10 bilayer) PAH-amaranth LBL films alongwith the pure amaranth solution, PAH-amaranth mixture solution and amaranth microcrystal spectra for comparison.

The pure amaranth solution fluorescence spectrum shows broad and intense 0-0 band with peak at 600 nm alongwith a weak hump at around 670 nm. The PAH-amaranth mixed (1:1 volume ratio) solution fluorescence spectrum shows intense 0-0 band at 630 nm and a broad and prominent band with peak at around 710 nm. The shifting of 0-0 band as well as development of longer wavelength band in the mixed solution spectrum was not readily explicable. However, a comparison with the fluorescence spectra of different layered LBL films as well as also with the microcrystal spectrum readily explain the origin of this band. Figure 4c also shows the fluorescence spectra of different layered LBL films as well as also the microcrystal spectrum. In all the cases the 0-0 band was observed to be at 625 nm. A shift of about 25 nm in comparison to the pure solution fluorescence spectrum. Moreover at higher number of layered LBL films and that in the microcrystal spectra, the longer wavelength broad band and the shift of 0-0 band of amaranth-PAH mixed solution spectrum is owing to the closer association of amaranth molecule while tagged onto the polymer backbone of PAH molecules and consequent formation of aggregates.

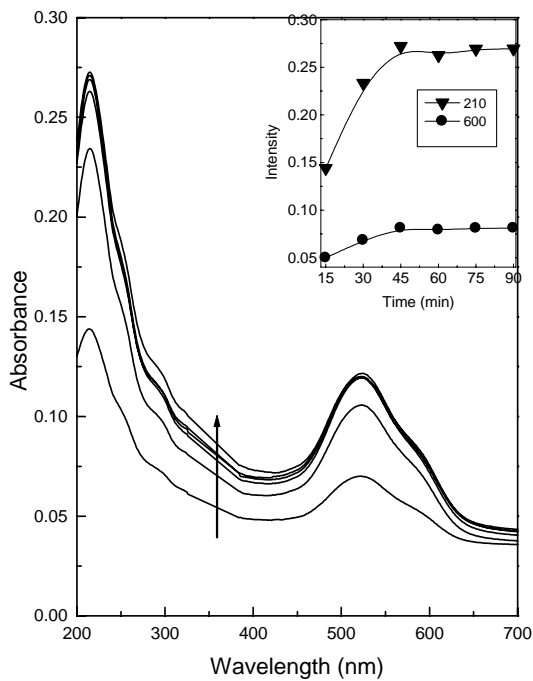
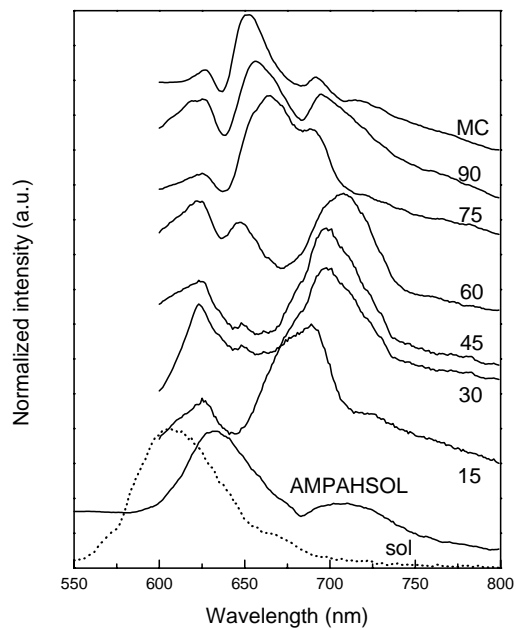

*Figure 5a:* Absorption spectra of 1-bilared amaranth-PAH layer-by-layer self assembled film with PAH deposition time 15 minute and different dye (amaranth) deposition times viz. 15, 30, 45, 60, 75 and 90 minutes. Inset shows the plot of intensities of 210 nm and 600 nm band as a function of time.

*Figure 5b:* Fluorescence spectra of 1-bilared amaranth-PAH layer-by-layer self assembled film with PAH deposition time 15 minute and different dye (amaranth) deposition times viz. 15, 30, 45, 60, 75 and 90 minutes alongwith the microcrystal spectrum (MC) amaranth-PAH mixture (1:1 volume ratio) solution (AMPAH) and pure amaranth solution spectrum (SOL).

Moreover the most interesting thing observed in the fluorescence spectra of different layered LBL films is that with changing layer number, the broad band system in the 650-750 nm region, as observed at lower number of layers, changes its intensity distribution among various bands within itself. Actually this broad band is an overlapping of various band systems, which become later prominent at higher number of layers. With the increase in number of layers (7, 8 and 9 bilayer) the fluorescence spectra give intense band at around 650 nm along with the 690 nm band and a weak hump at around 700 nm. At 10 bilayer LBL film the broad band at about 710 nm becomes prominent. This broad band is actually an overlapping of two bands one at 690 nm and a weak hump at around 710 nm. These two bands are prominent in microcrystal spectrum.

As discussed earlier the origin of this band system is solely due to the closer association of amaranth molecules and subsequent deformation of electronic levels in the molecular system.

The origin of this longer wavelength band with peak at 710 nm is solely due to the closer association and aggregate formation of amaranth molecules. However to check the nature of this aggregate and to confirm whether dimer or higher order n-meric species exist within this aggregate we have employed a traditional and conventional spectroscopic technique namely, excitation spectroscopic studies. Figure 4d shows the excitation spectra of 10 bilayer PAH-amaranth LBL film monitored at 625, 650 and 710 nm. From the figure it was observed that the excitation spectra monitored at 625 nm and 650 nm have almost similar band pattern. However, for 710 nm monitoring wavelength, the most interesting thing is that the excitation spectra is totally different than for those whose monitoring wavelengths are 625 and 650 nm. This certainly brings to the conclusion that there are some dimeric or higher order n-meric species exist in the LBL films. These different kinds of species predominate in the LBL films with increasing layer number.

The absorption spectra of PAH-amaranth 1-bilayer LBL films with different dipping time were shown in figure 5a. Here in all the cases the polycation (PAH) deposition time was kept fixed at 15 minutes but the dye (amaranth) deposition times were different, namely, 15, 30, 45, 60, 75 and 90 minutes. From the figure it was observed that the intensity of the absorption spectra increased for the films with dye deposition time up to 45 minute and remained almost constant for the dye deposition time greater than 45 minutes. This was also evidenced from the plot of the intensity of the absorption maxima versus time (inset of figure 5a). This indicate that the interaction of amaranth molecule with the PAH layer was completed within 45 minutes and after 45 minutes no PAH molecule remained free within the film for further interaction with the amaranth molecule.

Figure 5b shows the fluorescence spectra of PAH-amaranth 1-bilayer LBL films with different dye deposition time alongwith the amaranth solution, PAH-amaranth mixture solution and microcrystal fluorescence spectra for comparison.

From the figure it was observed that the fluorescence spectra of the films with dye deposition time up to 45 minute possess intense 0-0 band at 625 nm alongwith a broad band at 690 nm. But the fluorescence spectra with greater deposition time ($\geq$ 60 minute) were almost similar to the microcrystal fluorescence spectrum except a little change in intensity distribution among various bands. This may be due to the formation of low dimensional microcrystalline aggregates in the LBL films when the interaction gets completed.

**Conclusion:**

In conclusion our results show that self assembled films of low molecular weight dye amaranth and polycation Poly (allylamine hydrochloride) (PAH) can be prepared by electrostatic alternating Layer-by-Layer (LBL) adsorption process almost without any material loss. The UV-Vis absorption and fluorescence spectra of amaranth solution reveal that with the increase in amaranth concentration in solution the aggregated species start to dominate over the monomeric species. New aggregated band at 600 nm was observed in amaranth-PAH mixture solution absorption spectrum. A new interaction band at 710 nm in the amaranth-PAH mixture solution fluorescence spectrum was observed due to the closer association of amaranth molecule while tagged onto the polymer backbone of PAH and consequent formation of aggregates. The broad band system in the 650-750 nm region in the different layered florescence spectra changes in intensity distribution among various bands with changing layer number and at 10 bilayer LBL films the band at 710 nm becomes prominent. Dimeric or higher order n-meric species exist in the LBL films which were confirmed by excitation spectroscopic studies. Almost 45 minute was required to complete the interaction between amaranth and PAH molecules in the 1-bilayer LBL film.

**Acknowledgement:**

The authors are grateful to DST, UGC and CSIR Govt. of India for providing financial assistance through FIST-DST Project No. SR/FST/PSI-038/2002 and UGC minor research project No. Ref No. F.1-1/2000(FD-III)/399 and Sanction Ref. No.F. 31-30/2005 (SR) and CSIR project No. 03(1080)/06/EMR-II.